# How to "Censor" a Black Hole Singularity and Beyond

Asher Klatchko (klatchko@reed.edu)

April 29, 2016

**Abstract**

We argue that a black hole can be viewed as a gravitational optical element that images its interior onto the horizon. Being diffraction limited the *Airy hyper-ball* that forms as a result of interference of gravitational waves, does not allow for infinite resolution. Thereby effectively censoring the geodesics' singularity at $r \to 0$. The mechanism proposed must imply intrinsic angular momentum as in the Kerr black hole solution. However, assuming that a static black hole develops from an initial Kerr solution the destroyed hyper-ball is delegated to the horizon by the principle of increasing entropy. This lost information about the system's resolution is now packed on the horizon, which in effect provides the censorship. Should the mechanism be applicable to the early universe we suggest that the low variance circles seen in the CMB sky are traces of such an interference pattern.

**Introduction**

General relativity (GR) predicts both the existence of a black hole (BH) solution to Einstein's equation, and gravitational radiation. The direct detection of gravitational waves by LIGO [0], consistent with a simulation of the merging of two BHs each of ~30 $M_\odot$, suggests that (1) gravitational waves are a reality, and (2) that a collapse of a pair of BHs into a daughter BH is a dynamical system whose physics can be traced in a simulation. In fact it has been estimated that quite a large portion of the mass of the system has been lost during the merge to gravitational waves. Gravitational waves travel outwards from the source, carrying energy and information about the nature of the source. Following Feynman [1]: "*[If] energy is absorbed the wave must get weaker. How is this accomplished? Ordinarily through interference. To absorb, the absorber parts must move, and in moving generate a wave which interferes with the original wave in the so-called forward scattering direction, thus reducing the intensity for a subsequent absorber.*"

A central problem of the BH solution is its singularity. Unlike the chart singularity of the Schwarzschild metric the Eddington-Finkelstein coordinates [2] suggests a physical curvature singularity at $r = 0$, since $R^{abcd}R_{abcd} \propto 1/r^3$ as $r \to 0$. See for example "Gravitational Collapse: The Role of General Relativity", R. Penrose, [3]. For a careful analysis of the dependence of the singularity on $r$ see [4].

*So how does one "censor" the singularity*? As unique as the BH solution is, it is likely to evolve from a massive star (or maybe a lump of "dark matter" [5]), a process that can be described by the standard physics' "toolbox". In this note, following what Wheeler has called the "radically conservative" strategy, we will try to use the physics in the known "toolbox" to outline a possible mechanism for censorship of the BH singularity. We shall argue that the inability to resolve the caustic $r \to 0$ geodetic singularity suggests that its existence is irrelevant to the physics at hand and is clouded by some complementary aspect of the wave phenomena. In Berry's language – "ray cuastics are *complementary* to wave-front dislocations. For the following reasons. On a caustic, the intensity (in the shortwave limit) is infinite, whereas on a dislocation the intensity is zero." [6]. Much like the Heisenberg uncertainty relations in which context one can not talk about a precise location of a particle or its exact energy. In a black hole one is not able to resolve the singularity because the system is diffraction limited.

**Optics**

In GR, matter and spacetime are decoupled, that is, matter warps spacetime, which in turn constrains its movement, leading to the radiation of gravitational waves that ripple spacetime. In an accelerated expanding universe spacetime is generated copiously. In GR based cosmology this expansion is due to the cosmological constant acting as a source of negative pressure. What does happen to spacetime in a BH? Here we can only use metaphors to describe the geodesic behavior such as stretching, bending, falling into a singularity etc.

One way of describing the trapping of light in a BH is to think of it as an optical element with a high index of refraction. Light is slowed down by a factor $c/n$, and trapped due to the optical characteristics of the BH. However fast spacetime geodesics converge into the singularity, we will assume that information e.g. gravitational waves (the ripples of spacetime) can not move faster than the speed of light. For example consider the Painlevé–Gullstrand (PG) coordinates for a BH [7]. In the PG system a test particle inside the Schwarzschild horizon may fall faster than $c$ however information of its where about does not propagate faster than the speed of light.

In GR the relation between curvature of spacetime and the effective index of refraction can be calculated and has been measured a number of times since Eddington's celebrated measurement of the bending of light by the sun during the May 1919 eclipse [8]. As curvature increases so does the refractive index. Assuming that the index of refraction is bounded below by the weak field approximation $\sim 1 + \frac{2M}{r}$, [9] and taking into account the divergence near the 'singularity', $\sim \frac{1}{r^3}$, we shall formulate the refractive index as a cubic polynomial in powers of $\frac{M}{r}$. At small r ($r \ll 2M$), the index of refraction, $n(r)$, may grow to large values, effectively deeming the BH a strong optical element. As such we expect it to affect the gravitational waves at the BH interior. The resolution limit of this optical system is of the order of the characteristic wavelength divided by the numerical aperture $\sim \frac{\lambda}{n(r)\sin(\vartheta)}$. For the Kerr metric it can be shown that the numerical aperture is a function of the ergosurface outer and inner horizons(s), the mass and the angular momentum of the black hole only (see for example [10]):

$$(1) \quad sin^2(\vartheta) = 1 - \frac{4M^2 - (r_+^E - r_-^E)^2}{4(J/M)^2}.$$

Eq. (1) is consistent with the "no hair theorem" as it relates the optical properties of the black hole only to its fundamental attributes. A finite resolution would mean a finite numerical aperture or a finite collection angle of the gravitational waves. In what follows we will argue that the resolution cannot be infinitely small because this optical element is diffraction limited.

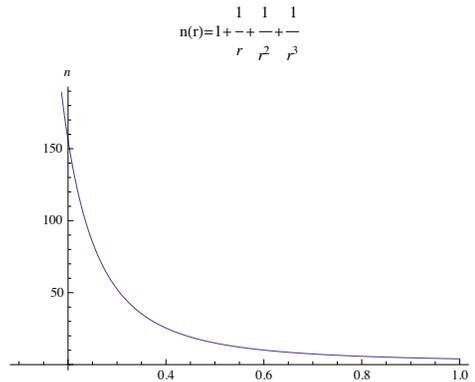

**Figure 1.** Index of refraction, *n(r)*, as a function of *r/2M*

Since geodesics emerge as solutions to the Euler-Lagrange equations, resulting from Fermat's principle in a curved space time, the vector field that governs the behavior of their paths is the *geodesic spray* of the Riemannian metric $G_{ij}(r)$, $G^i = \Gamma^i_{jk}(r) r'^j r'^k$, with, $\Gamma^i_{jk}(r)$, the Christoffel coefficients (eq. 1.3.10 in [12]). In this form, geodesic motion is described through a Riemannian spacetime with a meaning that is derived from Huygens' principle, which relates geodetic path to the motion of a wave front (see feature 10, in section 1.14 *Ten geometrical features of ray optics*, page 97 in [12]). In what follows we will assume that the Huygens principle holds for gravitational waves and investigate their interference as a way of avoiding the geodetics' singularity.

**Diffraction**

Imagine the collapse of a massive celestial object with a quadrupole moment generating gravitational waves at some characteristic spectrum, which could be strongly red shifted due to the strong field. The source is of some unknown spatial dimensions smaller or equal to the size of the horizon. Note that the emerging waves are expected to show a bunching effect due to the spatial coherence of the source. This is the famous Hanbury Brown and Twiss effect [11]. For simplicity assume that the waves can be described by plane-fronted waves with parallel rays or pp-waves for short [13][a]. As the waves propagate in this curved spacetime they interfere and form a diffraction pattern [13]. Recall that for a complete destructive interference of every one of the orders the phases have to add up to $\frac{(2n+1)}{2}\pi$, for every *n*. We argue that this is impossible. On information grounds alone, complete destructive interference would mean complete loss of information. However, should this happen then by the principle of increasing entropy the lost information would now be packed on the horizon area, suggesting that the horizon provides the necessary censorship.

---

[a] Note however that any other form of wave may lead to interference – see M. Berry in [6]

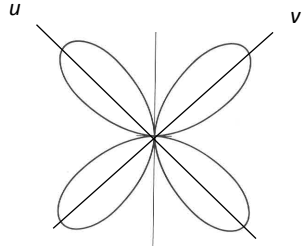

Quadrupole radiation pattern

**Figure 2.** The pattern of EM radiation from a quadrupole is used as template for gravitational waves

Since the waves come from a quadrupole source, qualitatively, we can approximate the source as a superposition of two orthogonal slits in the $u$ and $v$ direction (the orthogonal axes of the quadrupole radiation pattern above). At a distance, $L \gg \lambda$, the phase difference between the contributions from the two slits is equal to:

$$(2)\ \Delta\varphi = \frac{2\pi}{\lambda}((L^2 + u^2)^{\frac{1}{2}} - (L^2 + v^2)^{\frac{1}{2}}).$$

Upon using the binomial expansion, eq. (2) can be written as,

$$(3)\ u^2 - v^2 \approx \lambda L = constant.$$

The fringes are thus rectangular hyperbolae. We observe that although spacetime geodesics may crunch into a 'singularity' its ripples do not!

The ripples form a diffraction pattern whose fringes behave like rectangular hyperbolae as depicted in the picture below.

Rectangular Hyperbolae

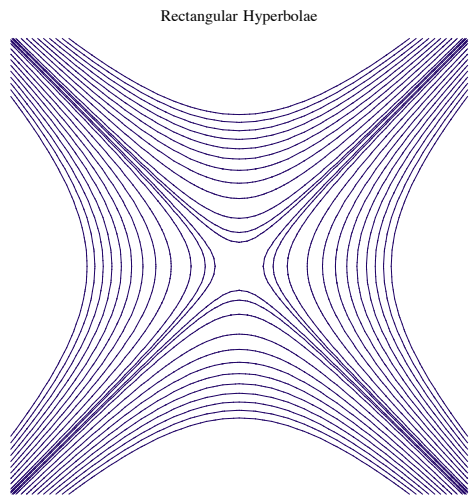

**Figure 3.** The diffraction pattern of two slits in the plane wave approximation

A pp-wave is a spacetime with a metric,

$$g = -2dudv - (f(u)(x^2 - y^2) + 2g(u)xy)du^2 + dx^2 + dy^2,$$

where $f(u)^2 + g(u)^2 \not\equiv 0$, and any $f$ and $g$ has vanishing Ricci tensor. It has been shown by Penrose that this mathematical setting leads to a lensing effect if the profile functions $f$ and $g$ are differentiable [15]. One often considers profile functions $f$ and $g$ with Dirac-delta-like singularities ("impulsive gravitational waves"). Nevertheless, rigorous and regularized solutions can be shown to exist which ensure distributional limits corresponding to physical expectations, suggesting that the geodesics correspond to refracted, broken straight lines [16]. The conclusion of theorem 2-5.9 of Ehlers and Kundt [17] is "[that] it is permissible to think of the graviton field independent of any matter by which it be generated. This corresponds to the existence of source-free photon fields in electro dynamics." As such it is enough to consider the addition of gravitational waves as in the Rayleigh-Sommerfeld diffraction integral and argue that diffraction must happen because it lowers the energy of the system. We note that the rectangular hyperbolic pattern is evident from the metric and on conformal mapping grounds may be a $z^2$ transform like of the axis of the quadrupole moment (see theorem 2-5.10 in [17]). On a qualitative basis we argue that the diffraction pattern is a focusing surface analogous to the Airy disk which forms in optics. In analogy with the Airy disk, the diffraction zone that is formed − an *Airy hyper-ball* − deems the BH diffraction limited. This 'Airy hyper-ball' does not allow for infinite resolution thereby censoring the 'singularity'.

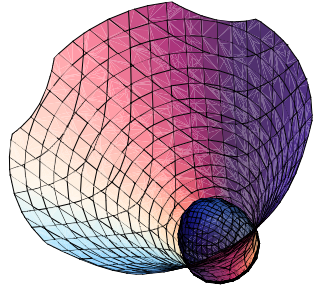

**Figure 4.** A depiction[b] of the Airy hyper-ball diffraction pattern at small $r$

The quadrupole source implies intrinsic angular momentum, which deems the BH a Kerr BH (KBH). In the Kerr spacetime the Ring singularity happens at,

$$(4) \quad x^2 + y^2 = a^2.$$

Since eq. (3) is always smaller than eq. (4),

$$(5) \quad x^2 - y^2 < x^2 + y^2 = a^2,$$

---

[b] We plot the Airy hyper-ball as the level sets of the Petzval invariant of the axisymmetric space $S^2 = X_1 X_2 - X_3^2$ (see section 1.4 in [13])

It appears that the diffraction zone is well located within the Ring. Moreover, we can estimate the wavelength's bounds for $L \leq r_-$:

$$(6) \quad \lambda_{\downarrow L \leq r_-} < \frac{a^2}{r_-} = \frac{a^2}{m - \sqrt{m^2 - a^2}},$$

where, $r_-$, is the inner event horizon and $a = J/m$. We note that as the angular momentum, $J$, is approaching its maximal value, $J_{max} \to m^2$, the wavelength grows as $\sqrt{J}$,

$$(7) \quad \lim_{J \to m^2} \lambda \propto m = \sqrt{J}.$$

On the other hand, as $J \to 0$ (the static Schwarzschild solution), the wavelength is infinitely shifted towards the infrared:

$$(8) \quad \lim_{J \to 0} \lambda = \frac{J^2}{m^2(m - \sqrt{m^2})} + O(J)^3.$$

Note that in the quantum regime $\lambda$ is of the order of the Compton wavelength, $(1/m)$, but grows as, $\sqrt{J} \sim m$, for large angular momenta, thereby approaching a constant. This suggests that the diffraction zone does not suffer from UV divergences.

**Steady state censorship**

A KBH may lose rotational energy via its ergosphere and become a static Schwarzschild BH. At this point the Airy hyper-ball is destroyed and the information about the system's resolution is delegated to the horizon by increasingly larger wavelengths. This suggests that censorship starts under the conditions of a rotational steady state but may decay into a static state. There exists ample evidence for astrophysical jets, to provide the necessary astronomical fingerprints for a steady state censorship of KBH (see [18] and references therein). But by this argument even static BH would not appear naked. For the various orders of interference to stay stable there has to be a force balancing the interaction of the waves' self-energy that may lead to implosion [19]. It is plausible that gravitomagnetic effects prevent the implosion by balancing off the gravitational self-energy of the formed front with its rotational energy. Alternatively, one could describe the stability in the language of critical phenomena as a result of scaling and self similarity which is beyond the scope of this note [20].

**Gravitomagnetism and the Aharonov Bohm effect**

It has been pointed out that an analog of the Aharonov Bohm effect (AB) happens in correlation with intrinsic angular momentum of a spinning body (see for example [21] and references therein). The gravitomagnetic potential introduces a nonlocal effect as in the AB case for electrodynamics. Assuming that the said non-locality happens in the strong field too, leading to another source of information to reckon with. The extra phase due to the BH's angular momentum contributes to the formation of the diffraction pattern. This adds a nonlocal, aspect to the censorship mechanism.

**Cosmic censorship and low-variance circles in the CMB sky**

Should the mechanism outlined in this note be applicable to the early universe it is possible that the existence of low variance circles in the cosmic microwave background

(CMB) sky, as first claimed by Gurzadyan and Penrose [22], are traces of the diffraction pattern described above, rather than a result of conformal cyclic cosmology (CCC) [23]. Gurzadyan and Penrose (http://arxiv.org/pdf/1011.3706v1.pdf).

Indeed for the inflationary phase that takes over soon after the universe emanates from the initial singularity the far field approximation of pp-waves is rather adequate. The emerging pattern may manifest the "Evolution from a Fraunhofer to a Pearcey diffraction pattern" [24]. It is still necessary to work out the details of this censorship mechanism in order to predict the outcome in terms of the pattern expected or fit the empirical evidence to the mechanism.

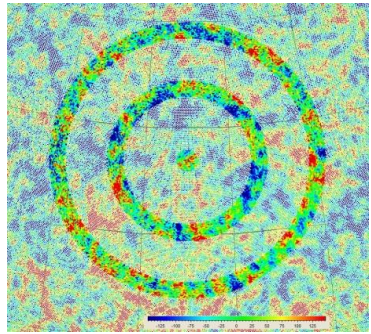

**Figure 5.** An example of low variance circles in WMAP data of recorded CMB from

**Discussion and Summary**

We have argued that either by the formation of a diffraction pattern, which we have termed the *Airy hyper-ball*, or by *complete destruction of the interference pattern on each and every order* a censorship mechanism exists which obstructs infinite resolution of the whereabouts of the black hole singularity. Should the pattern be destroyed, then by the Bekenstein-Hawking entropy principle, $S_{BH} \propto Area$, the compensation for the loss of information proceeds via an increase of the horizon area. We note that either scenario supports the holographic view of the horizon as a gravitationally formed image of the black hole interior. In optical terms this relation would be called imaging. A result which is consistent with the emergent 'holographic' nature of gravity first proposed by Gerard 't Hooft.

Since the proposed censorship mechanism necessitates intrinsic angular momentum it cannot censor the Schwarzschild static black hole singularity unless the static hole is a decay byproduct of a Kerr hole. This is not a far-fetched idea as the Schwarzschild solution is the limit of the Kerr metric when the angular momentum vanishes, $J \to 0$. The fact that a singularity is not resolved is synonymous to the Heisenberg uncertainty principle, which doesn't permit simultaneous knowledge about the exact state of a system with infinite resolution. Here the problem of infinite tidal forces at the $r \to 0$ singularity turns out to be moot because the singularity cannot be resolved. We point out the possibility that the interference pattern may explain the non-random concentric sets of low-variance circular rings in the WMAP data that were formerly claimed as an indication for CCC.

**References**

[0] B. P. Abbott et al. (LIGO Scientific Collaboration and Virgo Collaboration) Phys. Rev. Lett. 116, 061102
http://dcc.ligo.org/public/0122/P1500229/028/GW150914_burst.pdf
[1] R.P. Feynman, remarks made in the 1957 Chapel Hill conference on The Role of Gravitation in Physics, http://www.edition-open-access.de/sources/5/34/index.html
[2] A. S. Eddington, Nature, 113, 192 (1924), D. Finkelstein, Phys. Rev., 110, 965 (1958)
[3] R. Penrose, General Relativity and Gravitation, Vol. 34, No. 7, July 2002
[4] T. Kawai and E. Sakane, http://arxiv.org/abs/gr-qc/9707029
[5] Bird et al., http://arxiv.org/abs/1603.00464
[6] M. Berry, Les Houches, Session XXXV, 1980 – Physics of Defects, North-Holland Publishing Company, 1981, https://michaelberryphysics.files.wordpress.com/2013/07/berry105.pdf
[7] P. Painlevé, C. R. Acad. Sci. (Paris) 173, 677-680 (1921), A. Gullstrand, Mat. Astron. Fys. 16(8), 1-15 (1922)
[8] F.W. Dyson, A.S. Eddington and C. Davidson, http://rsta.royalsocietypublishing.org/content/roypta/220/571-581/291.full.pdf
[9] P. Boonserm et al., http://arxiv.org/abs/gr-qc/0411034
[10] M. Visser, "The Kerr spacetime: A brief introduction", http://arxiv.org/abs/0706.0622
[11] R. Hanbury Brown and R. Q. Twiss, Nature 177, 27 (1956)
[12] D. D. Holm, "Geometric Mechanics", 2$^{nd}$ Edition, http://wwwf.imperial.ac.uk/~dholm/classnotes/GeomMech1-2nd.pdf
[13] H.W. Brinkmann, "Einstein spaces which are mapped conformally on each other", Math. Ann., 94, 119–145, (1925).
[14] V. H. Schultheiss et al. "Optics in Curved Space", PRL 105,143901 (2010), D. Bar, "Gravitational holography and trapped surfaces" and references therein (http://arxiv.org/abs/gr-qc/0605115v2)
[15] V. Perlick "Gravitational Lensing from a Spacetime Perspective", Living Rev. Relativity 7 (2004), 9, http://relativity.livingreviews.org/Articles/lrr-2004-9/
[16] M. Kunzinger, R. Steinbauer, http://arxiv.org/abs/gr-qc/9806009
[17] Ehlers, J., and Kundt, W., "Exact solutions of gravitational field equations", in Witten, L., ed., Gravitation: An Introduction to Current Research, pp. 49–101, (Wiley, New York, 1962).
[18] R. K. Williams, The Astrophysical Journal, 611, 952-963 (2004)
[19] A. M. Abrahams, C. R. Evans, Phys. Rev. D 46 10 (1992)
[20] C. Gundlach, "Critical Phenomena in Gravitational Collapse", Living Rev. Relativity 2, (1999), 4. URL http://www.livingreviews.org/lrr-1999-4
[21] J.G. de Assis, C. Furtado and V.B. Bezerra, Gravitation & Cosmology, Vol. 10 (2004), No. 4 (40), pp. 295–299
[22] V.G. Gurzadyan, R. Penrose, "More on the low variance circles in CMB sky" http://arxiv.org/abs/1012.1486

[23] R. Penrose, "Cycles of Time: An Extraordinary New View of the Universe." (Bodley Head, London, 2010), ISBN 978-0-224-08036-1.
[24] J. F. Nye, J. Opt. A: Pure Appl. Opt. 5 (2003) 495–502, http://iopscience.iop.org/article/10.1088/1464-4258/5/5/310/pdf